\documentclass[%
 preprint,
superscriptaddress,
 amsmath,amssymb,
 aps,
 prl,
]{revtex4-2}

\usepackage{graphicx}
\usepackage{dcolumn}
\usepackage{bm}
\usepackage{color}
\usepackage[colorlinks=true,urlcolor=blue,citecolor=blue,linkcolor=blue,bookmarks=false,pdfstartview={FitH}]{hyperref}
\usepackage{ulem}
\usepackage{amssymb}

\newcommand{\LST}{LaSb\textsubscript{x}Te\textsubscript{2-x}}
\newcommand{\oLST}{\textit{o}-LaSbTe}
\newcommand{\tLST}{\textit{t}-LaSbTe}
\begin{document}
    
\title{Electronic switching of topology in LaSbTe}

\author{J. Bannies}
\email{jbannies@chem.ubc.ca}
\affiliation{Quantum Matter Institute, University of British Columbia, Vancouver, British Columbia, V6T 1Z4, Canada}
\affiliation{Department of Chemistry, University of British Columbia, Vancouver, Canada V6T 1Z1, Canada}

\author{M. Michiardi}
\affiliation{Quantum Matter Institute, University of British Columbia, Vancouver, British Columbia, V6T 1Z4, Canada}
\affiliation{Department of Physics \& Astronomy, University of British Columbia, Vancouver, British Columbia, V6T 1Z1, Canada}

\author{H.-H. Kung}
\affiliation{Quantum Matter Institute, University of British Columbia, Vancouver, British Columbia, V6T 1Z4, Canada}
\affiliation{Department of Physics \& Astronomy, University of British Columbia, Vancouver, British Columbia, V6T 1Z1, Canada}

\author{S. Godin}
\affiliation{Quantum Matter Institute, University of British Columbia, Vancouver, British Columbia, V6T 1Z4, Canada}
\affiliation{Department of Physics \& Astronomy, University of British Columbia, Vancouver, British Columbia, V6T 1Z1, Canada}

\author{J. W. Simonson}
\affiliation{Department of Physics, Farmingdale State College, Farmingdale, NY 11735, USA}

\author{M. Oudah}
\affiliation{Quantum Matter Institute, University of British Columbia, Vancouver, British Columbia, V6T 1Z4, Canada}

\author{M. Zonno}
\altaffiliation[Present address: ]{Synchrotron SOLEIL, Saint-Aubin 91192, France}
\affiliation{Canadian Light Source Inc., Saskatoon, Saskatchewan, S7N 2V3, Canada}

\author{S. Gorovikov}
\affiliation{Canadian Light Source Inc., Saskatoon, Saskatchewan, S7N 2V3, Canada}

\author{S. Zhdanovich}
\affiliation{Quantum Matter Institute, University of British Columbia, Vancouver, British Columbia, V6T 1Z4, Canada}
\affiliation{Department of Physics \& Astronomy, University of British Columbia, Vancouver, British Columbia, V6T 1Z1, Canada}

\author{I. S. Elfimov}
\affiliation{Quantum Matter Institute, University of British Columbia, Vancouver, British Columbia, V6T 1Z4, Canada}
\affiliation{Department of Physics \& Astronomy, University of British Columbia, Vancouver, British Columbia, V6T 1Z1, Canada}

\author{A. Damascelli}
\email{damascelli@physics.ubc.ca}
\affiliation{Quantum Matter Institute, University of British Columbia, Vancouver, British Columbia, V6T 1Z4, Canada}
\affiliation{Department of Physics \& Astronomy, University of British Columbia, Vancouver, British Columbia, V6T 1Z1, Canada}

\author{M. C. Aronson}
\email{meigan.aronson@ubc.ca}
\affiliation{Quantum Matter Institute, University of British Columbia, Vancouver, British Columbia, V6T 1Z4, Canada}
\affiliation{Department of Physics \& Astronomy, University of British Columbia, Vancouver, British Columbia, V6T 1Z1, Canada}

\date{\today}
\begin{abstract}
\textbf{In the past two decades, various classes of topological materials have been discovered, yet the deliberate control of topology in a single material remains largely unexplored. Here, we demonstrate full experimental control over the topological nodal loop in the square-net material \LST\ by chemical substitution and electron doping. Using angle-resolved photoemission spectroscopy (ARPES), we show that changing the antimony concentration x from 0.9 to 1.0 in the bulk opens a gap larger than 400~meV in the nodal loop. Symmetry analysis establishes that this effect originates from the breaking of \textit{n} glide symmetry in the square-net layer. Remarkably, the same topological phase transition can also be driven reversibly on the surface of \LST\ by \textit{in situ} chemical gating via potassium deposition, enabling on-demand switching of topology. The control parameter for both the bulk and surface transition is the electron concentration, providing a pathway towards applications based on switching topology by electrostatic gating.}
\end{abstract}

\maketitle

\section{\label{sec:intro}Introduction}
\noindent Symmetry is key to understanding the low-energy electronic structure of topological materials, which are generally characterized by a non-zero topological invariant. 
The existence of Dirac fermions in condensed matter systems, for example, is protected by time-reversal symmetry~\cite{Zhang2009a,Hasan2010a} or crystalline symmetry~\cite{Ando2015a}. Many of the phenomena found in topological materials are related to the Dirac fermions near the Fermi level, such as the quantum spin Hall effect in two-dimensional topological insulators~\cite{Bernevig2006a,Koenig2007a,Wu2018a}, quantum anomalous Hall effect in magnetically-doped topological insulators~\cite{Chang2013a}, or extremely large magnetoresistance in Dirac semimetals~\cite{Liang2014a}. To harness these effects for device applications, it is crucial to control the topology of a material, rather than simply observing it. 

In the most fundamental case, one can imagine driving a topological phase transition from the topologically trivial state to the non-trivial state by external input. 
To achieve such a topological phase transition, it is desirable to find a way to tune the symmetry that protects the topological phase. Successful attempts to destabilize the topological phase include the gap opening at the surface Dirac cone in topological insulators by breaking of time-reversal symmetry~\cite{Chang2013a,Chen2010a,Zhang2013a}, and in topological crystalline insulators by breaking mirror symmetry on the surface~\cite{Zeljkovic2015a}. 
Similarly, topological phase transitions were achieved in topological insulators by continuously tuning the strength of spin-orbit coupling via bulk chemical substitution~\cite{Xu2011a,Xu2012a,Wojek2015a}.
However, all these realizations of topological phase transitions rely on bulk chemical substitution, which is an irreversible process that cannot be triggered and controlled by external input.

To fully exploit the considerable promises of future technologies based on topological materials, it is important to find a fast and reliable switching knob that can turn the topological states on and off. To be in accord with current state-of-the-art, an electronically-driven process is the most desirable since it can be controlled by fast switching of electric fields. Such a process was demonstrated in quantum-confined Na$_3$Bi~\cite{Collins2018a}; however, the extreme air sensitivity of Na$_3$Bi is unfavorable for applications. 

Here, we report the electronically-driven opening of a large gap at the Fermi level in the Dirac nodal loop of the square-net material \LST. Increasing x from 0.9 to 1 by bulk chemical substitution leads to a structural phase transition that breaks the \textit{n} glide symmetry of the square-net layer by distortion into zig-zag chains. Combining angle-resolved photoemission spectroscopy (ARPES) with tight binding simulations, we establish that this symmetry breaking triggers 
the opening of a large topological gap concomitant with a topological phase transition. The control over this topological gap distinguishes our study from a previous study that focused on a symmetry-enforced nodal line at the Brillouin zone boundary~\cite{Wang2021a}.
Importantly, we can also realize the opening and closing of the gap \textit{in situ} by chemical gating of the surface of \LST. 
The electronic nature of the transition opens the possibility of straightforward control of the nodal loop topology by electrostatic gating in devices based on \LST\ thin films. 

\section{\label{secr:main}Main}
\begin{figure*}[ht!]
	\centering
	\includegraphics[width=0.5\linewidth]{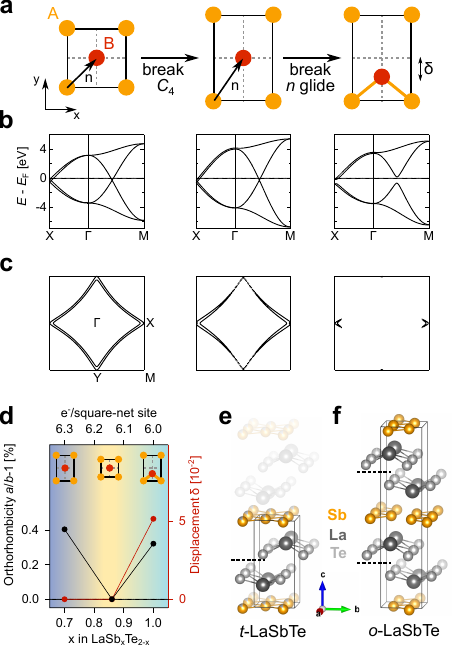}
	\caption{\label{fig1:square} \textbf{Gap opening of the square-net-derived nodal loop by symmetry lowering. }
	\textbf{a} Minimal model for a square-net with two sublattices A and B that is undergoing symmetry lowering from an ideal square-net to zig-zag chains by consecutive breaking of \textit{C}\textsubscript{4} and \textit{n} glide symmetry. 
    SOC is not included.
    \textbf{b} Breaking of \textit{C}\textsubscript{4} symmetry causes minor changes to the band structure along X-$\Gamma$-M and leaves the nodal loop gapless. Only breaking of \textit{n} glide symmetry causes the opening of a large gap in the nodal loop. 
    \textbf{c} Similarly, the diamond shaped Fermi surface of the ideal square-net experiences only minor warping by removal of \textit{C}\textsubscript{4} symmetry, whereas subsequent breaking of \textit{n} glide symmetry causes a major reduction in Fermi surface volume by gapping out large parts of the Fermi surface. 
    \textbf{d} All three stages of the model are experimentally realized in \LST\ as a function of antimony concentration x. The values for orthorhombicity and displacement, as determined from single crystal XRD, indicate breaking of \textit{C}\textsubscript{4} and \textit{n} glide symmetry, respectively. The partial substitution of antimony with tellurium for $x<1$ effectively increases the electron filling of the square-net layer. Error bars are smaller than the symbol size.
    \textbf{e,f} Crystal structures of \tLST\ and \oLST, corresponding to x=0.9 and 1, where the breaking of antimony square-nets into zig zag chains switches the topology, driving the transition from gapless to gapped nodal loop. These antimony layers lie below the surface after cleavage along the dashed lines.}
\end{figure*}

\noindent To understand which symmetries need to be broken to lift the topological protection of the nodal loop, we consider a simple model consisting of a square-net with a body-centered two-dimensional square lattice as shown in Fig.~\ref{fig1:square}a.
It can be described by two equivalent sublattices A and B, which are related by non-symmorphic \textit{n} glide symmetry - a mirror operation $M_\mathrm{z}$ followed by a (1/2, 1/2) lattice translation. 
We build a minimal tight binding model of this square lattice that uses only the \textit{p}\textsubscript{x,y} orbitals,
similar to the model used in Ref.~\cite{Deng2022a}. The effect of spin-orbit coupling (SOC) is not included in this model but will be considered later.
Distances and electron fillings were chosen to most closely represent the system under study in this work. It is known that the body-centering of the unit cell imposes a band folding that gives rise to degenerate band crossings~\cite{Klemenz2019a} as visible in Fig.~\ref{fig1:square}b.
We study the effect of symmetry lowering in this ideal square-net (Fig.~\ref{fig1:square}a) in two steps, by first removing fourfold symmetry and then \textit{n} glide symmetry, and track the changes in the band structure as well as the Fermi surface (Figs.~\ref{fig1:square}b,c). 

The band structure of the undistorted square-net is dominated by linear band crossings close to the Fermi level. These crossings are part of a nodal loop that spans the entire diamond-shaped Fermi surface and is protected by \textit{n} glide symmetry. 
The breaking of fourfold symmetry only has minor effects on the band structure, and importantly the nodal loop remains gapless because \textit{n} glide symmetry still translates atom A into atom B.
However, subsequent displacement of atom B from the center of the unit cell breaks \textit{n} glide symmetry, forming zig-zag chains and making the sublattices inequivalent. This symmetry lowering lifts the degeneracy and creates a gap in the nodal loop. 
The size of the gap depends on the degree of displacement of atom B; for the experimental value of the system under study, we obtain a large gap of several hundred meV.   
Correspondingly, the diamond-shaped Fermi surface disappears and only small electron pockets survive at the Brillouin zone boundary. Importantly, these pockets are formed from parabolic bands, in contrast to the linear bands of the original square-net.
Our minimal model highlights the requirement of breaking \textit{n} glide symmetry as the mechanism to gap the nodal loop, and also shows the resulting drastic change in the Fermi surface. It furthermore suggests that a change in crystalline symmetry between these extreme cases of square-net and zig zag chains causes the massless Dirac fermions to acquire a finite mass term. 
We note that the inclusion of SOC leads to a small gap in the nodal loop and turns the square-net layer into a quantum spin Hall insulator, even with preserved \textit{n} glide symmetry~\cite{Deng2022a}.
However, breaking this symmetry increases the gap by more than one order of magnitude and thus is a much stronger perturbation to the electronic structure than SOC, and in turn more suitable for applications. Yet, breaking the \textit{n} glide symmetry on-demand experimentally is not a trivial feat in square-net materials, as it requires selectively displacing the center atom of the unit cell.

To this end, we consider the curious case of the RESb\textsubscript{x}Te\textsubscript{2-x} (RE=rare-earth) systems, which feature a square-net of antimony atoms and are known for their rich phase diagrams~\cite{Dimasi1996a,Singha2021a,Lei2019a,Pandey2022a}, including a tetragonal to orthorhombic transition as a function of chemical substitution of antimony by tellurium ~\cite{Singha2021a,Lei2019a,Pandey2022a,Qian2020a}. We choose the specific case of \LST\ as the simplest system that is non-magnetic and is not subject to electronic correlation effects.
A previous study predicted a topological phase transition between two distinct phases of LaSbTe~\cite{Qian2020a}; however, it remained unclear how this transition can be realized in experiments, and the topological character was shown experimentally only for one of these phases based on the observation of nodal lines at the Brillouin zone boundary~\cite{Wang2021a}.

We grew single crystals of \LST\ and determined their structure and composition by single crystal X-ray diffraction (XRD) and wavelength dispersive X-ray spectroscopy.
Remarkably, we find that the square-net layer in \LST\ undergoes both symmetry lowering transitions required to realize our minimal square-net model, by simply changing the antimony concentration x, which is equivalent to changing the electron filling of the square-net. Our single crystal XRD data evince the selective breaking of fourfold and \textit{n}~glide symmetry in a narrow window of $0.7<\mathrm{x}<1$, as shown in Fig.~\ref{fig1:square}d, resulting in two distinct orthorhombic phases in close vicinity to an intermediate tetragonal phase.

Here, we focus on two neighboring phases of \LST: tetragonal LaSb\textsubscript{0.86}Te\textsubscript{1.14} with a gapless nodal loop, and orthorhombic LaSbTe with a gapped nodal loop. Single-crystal structure refinements (see SM) show that the former is characterized by an ideal antimony square-net layer with \textit{n} glide symmetry, whereas the latter features antimony zig-zag chains, i.e. broken \textit{n} glide symmetry, in the antimony layer  (Fig.~\ref{fig1:square}e,f). 
While an earlier report found the same orthorhombic structure for \LST\ with $x=1$~\cite{Dimasi1996a}, a more recent study found a tetragonal structure for that composition~\cite{Wang2021a}. These differences are most likely stemming from different growth conditions and slightly different resulting compositions.
From here on, we refer to the materials 
in our study
as \tLST~
($x=0.86$) and \oLST~($x=1$). We probe the band structure of the two phases 
by employing ARPES on single crystals of both \oLST\ and \tLST. 

\begin{figure*}[ht!]
	\centering
	\includegraphics[width=1\linewidth]{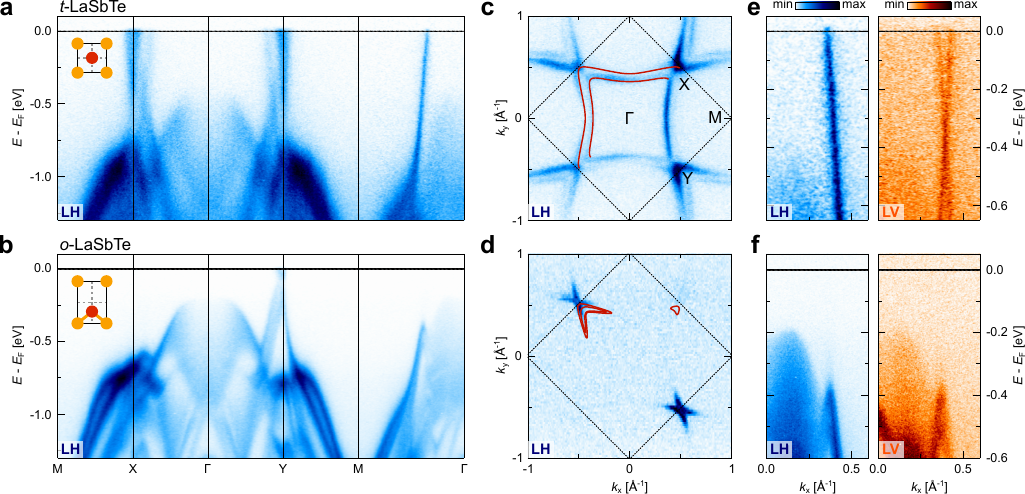}
	\caption{\label{fig2:bulk} \textbf{Topological phase transition from gapless to gapped nodal loop in \LST.}
    \textbf{a, b} ARPES spectra of \tLST\ and \oLST, respectively, along a high symmetry path. Bands forming the nodal loop are gapless and cross the Fermi level along $\Gamma$-M and $\Gamma$-X/Y in \tLST. Breaking of \textit{n} glide symmetry in \oLST\ causes these bands to open a massive gap at the X point and along $\Gamma$-M.
    \textbf{c, d} Fermi surfaces of \tLST\ and \oLST; the two concentric diamond shaped sheets in \tLST\ gap out almost entirely upon breaking of \textit{n} glide symmetry in \oLST, in excellent agreement with DFT calculations (red lines overlaid). The black dashed line indicates the Brillouin zone. 
    \textbf{e} Zoom-in spectra along $\Gamma$-M for \tLST\ acquired with linear horizontal (LH) and linear vertical (LV) polarization to probe matrix element effects (note that all data in panels \textbf{a}-\textbf{d} were measured with LH polarization). Both branches of the gapless nodal loop are visible 
    and the small gap expected from DFT calculations with SOC is not observed.
    \textbf{f} Same as panel \textbf{e} but for \oLST; with both LH and LV polarization we observed a $>$400~meV gap in the nodal loop along $\Gamma$-M.
    }
\end{figure*}

In Fig.~\ref{fig2:bulk}a,b we present an overview of the band structures of \tLST\ and \oLST\ measured over a wide energy window. 
For \tLST, linear bands cross the Fermi level along $\Gamma$-X and $\Gamma$-M. As expected from our minimal model, those bands are gapped by several hundred meV at X and along $\Gamma$-M in \oLST\ due to the broken \textit{n} glide symmetry. These differences are also reflected in the Fermi surfaces of the two phases, see Fig.~\ref{fig2:bulk}c,d. In \tLST, the linear bands form two concentric diamond-shaped 
and nested
Fermi surface sheets, whereas the Fermi surface of \oLST\ only consists of two small electron pockets, which are no longer nested.
These observations demonstrate the experimental realization of our square-net model and of the topological phase transition theoretically predicted in Ref.~\cite{Qian2020a}, and are in excellent agreement with complementary band structure calculations based on density functional theory (DFT).

Furthermore, we observe a strong dependence of the intensity of the nodal loop bands on the polarization of the incident light for \tLST: linear horizontal (LH) and linear vertical (LV) polarizations couple to opposite branches of the nodal loop, see Fig.~\ref{fig2:bulk}e.
The observed bands are gapless and linear with high Fermi velocities of 8.8(1)~eV$\cdot$\AA\textsuperscript{-1}, in agreement with previous reports~\cite{Wang2021a}. In contrast, the nodal loop is gapped by at least 400~meV along $\Gamma$-M in \oLST, regardless of the light polarization, as seen in Fig.~\ref{fig2:bulk}f. We define this lower bound by the valence band maximum energy with respect to the Fermi level, as we cannot probe the unoccupied conduction band states and therefore the conductivity gap directly in ARPES experiments. In addition, we observe a splitting of the valence band, which we attribute to the presence of two antimony layers per unit cell in \oLST\ (Fig.~\ref{fig1:square}f). A similar bilayer splitting was found in related square-net materials such as rare-earth tritellurides~\cite{Brouet2004a,Brouet2008a}. 

The opening of a gap by changing the crystalline symmetry of \LST\ via small variation in composition is the direct experimental demonstration that \textit{n} glide symmetry indeed protects the existence of the nodal loop in ideal square-net materials 
in the absence of SOC. When included, SOC opens a small gap of 40-50~meV even in \tLST, according to our DFT calculations~(see SM). However, we find no indications for such a gap in our ARPES spectra (Fig.~\ref{fig2:bulk}e). While the gap likely exists, its small size seems to be obfuscated by resolution and intrinsic $k_\mathrm{z}$ broadening. We note that the nodal loop reported here is more suited for manipulation than the genuine gapless nodal line previously reported~\cite{Wang2021a}, since this nodal line is gapless in both \tLST\ and \oLST\, even in the presence of SOC, as shown by symmetry analysis performed in~\cite{Leonhardt2021a,Hirschmann2021a}.
A complete description of the topology in the presence of SOC is offered by the symmetry indicators $\mathbb{Z}_{2,2,2,4}$. In this scheme, \tLST\ is a weak topological insulator with $\mathbb{Z}_{2,2,2,4}=(0010)$ and \oLST\ is topological crystalline insulator with $\mathbb{Z}_{2,2,2,4}=(0002)$~\cite{Qian2020a,Deng2022a}. The change in symmetry indicators signals a topological phase transition even in the presence of SOC, and describes the evolution of the overall topology of the material.

The doping dependence of the phase transition presents itself as a unique opportunity to study the origin of this transition. Because small changes in composition induce a phase transition, we can try to understand whether it is caused by the chemical pressure effect induced by chemical substitution of antimony by tellurium, or rather by the effective electron doping accompanying the substitution. If the transition is electronically-driven, it will allow for much easier and faster switching between the phases, such as by gating the material. 
Here, we demonstrate that the small tellurium excess needed to stabilize \tLST\ is equivalent to electron-doping the square-net layer. We emulate this property of the bulk material using \textit{in situ} chemical gating by stepwise potassium deposition to electron-dope the surface of \oLST.
We track the evolution of the electronic structure across the transition by measuring the Fermi surface and the $\Gamma$-M cut at each deposition step, as shown in Fig.~\ref{fig3:kdepos}a,~b. 

In the initial stages of the potassium deposition, the valence bands shift down in energy and the Fermi surface volume increases, indicating a rigid band shift of the Fermi level caused by the electron doping. However, starting from step 3, a qualitatively different behavior sets in. The Fermi surface undergoes a gradual redistribution of spectral weight and representative energy distribution curves (Fig.~\ref{fig3:kdepos}c) show the appearance of in-gap states in the nodal loop gap. Increasing the doping leads the nodal loop gap to collapse completely and the Fermi surface becomes fourfold symmetric, in stark contrast to the twofold symmetric Fermi surface of pristine \oLST. In addition, the bilayer splitting, characteristic of broken \textit{n} glide symmetry, disappears as the in-gap states fill the entire gap (see Fig.~\ref{fig3:kdepos}d).  
Effectively, the band structure of chemically gated \oLST\ resembles that of bulk \tLST, albeit with slightly lower electron doping as the nodal loop crossing lies about 90~meV higher in energy compared to \tLST.
The doping-controlled electronic phase transition of the \oLST\ surface is therefore the analog of the bulk electronic phase transition between \oLST\ and \tLST, for which electron doping is achieved by decreasing the antimony concentration x. 
Given the clear correspondence between broken \textit{n} glide symmetry and gapped nodal loop that we established in bulk \LST, it is likely that the surface crystal structure will follow the change in electronic structure even in our surface doping experiments, i.~e. restore \textit{n} glide symmetry in the surface antimony layer.
Remarkably, the \textit{in situ} doping-driven phase transition is shown to be completely reversible by desorbing the potassium from the sample surface (see Methods). As shown in Fig.~\ref{fig3:kdepos}e, the system transitions continuously back into the pristine surface of \oLST, breaking the \textit{n} glide symmetry and restoring the gap in the nodal loop.

Our observation of the electronic phase transition through both bulk chemical substitution and surface potassium deposition indicates that the main control parameter is electron doping; however, concomitant additional effects from introducing a new chemical species cannot be excluded even in the case of potassium deposition.
To validate carrier doping as the main control parameter for the electronic phase transition, we performed DFT calculations of two kinds: pure electron doping of bulk \oLST, and potassium adatoms on slab structures of \oLST. In the bulk, we relaxed the structure of \oLST\ for different amounts of electron doping and tracked the displacement of the center antimony atom as a function of doping, as shown in Fig.~\ref{fig4:dft}a. Upon the doping of about 0.13 electrons per formula unit, the displacement decreases to 0, equivalent to restoration of \textit{n} glide symmetry. Thus electron doping alone can explain the structural phase transition from \oLST\ to \tLST\ in bulk, highlighting the electronically driven nature of the phase transition without invoking the chemical pressure associated with the Sb-Te substitution. 
For the slab calculations, we relaxed 5-unit cell slabs of \oLST\ with one potassium atom per surface unit cell in two configurations (see SM for details). We find that the first antimony layer recovers \textit{n} glide symmetry, while deeper layers remain mostly unaffected, regardless of the exact position of the potassium on the surface (Fig.~\ref{fig4:dft}b). This result reproduces our experimental observation and supports our expectation that the structure is likely to follow the electronic transition of the surface layer. Further structural studies are needed to experimentally confirm whether the surface electronic phase transition is accompanied by the same structural distortion observed in bulk.

The surface electronic phase transition is characterized by a complex band structure change in the intermediate phase that encodes the microscopic mechanism underlying such behavior. To understand this change better, we probe the polarization dependence by using LH and LV polarizations similar to bulk \tLST\ and \oLST. The resulting band dispersions are presented in Fig.~\ref{fig5:cdw}a. 

In dosing steps 4 and 5, the in-gap states exhibit Dirac-like dispersion even though the nodal loop still exhibits a gap. These Dirac in-gap states move down in energy upon increasing electron doping; concomitantly, the bandwidth of the in-gap states increases until they span the entire gap in the fully dosed state in step~6. 
Our model (Fig.~\ref{fig1:square}a) provides no mechanism for in-gap states within the nodal loop gap. We conclude that the phase transition is accompanied by a more complex behavior than the simple simultaneous displacement of the central antimony atoms. The experimental observation of band folding in the in-gap states suggests an additional transition involving a charge density wave (CDW) phase. Indeed, CDW instabilities are commonly found in square-net materials.~\cite{Brouet2004a,Brouet2008a,Dimasi1996a,Lei2019a,Shin2005a,Malliakas2006a}. 

For a qualitative description of the observed behavior, we construct model CDWs using only antimony atoms in a supercell approach and calculate their electronic structure in the framework of DFT. Starting from the unit cell of a perfect square-net, we introduce a sinusoidal modulation of the center antimony's \textit{y} coordinate in each subcell, as illustrated for a 7x1x1 structure in Fig.~\ref{fig5:cdw}b. For simplicity, we focus on the unfolded band structure along $\Gamma$-M for several supercells of different size.  
For the 11x1x1 supercell, the Dirac cone characteristic for \tLST\ is broken up by several gaps such that the bandwidth of the subbands around the Dirac cone is reduced compared to the unperturbed Dirac cone (see Fig.~\ref{fig5:cdw}c). The band folding at the edges of the subbands is a hallmark of the CDW and resembles the band folding of the in-gap states in our experimental data.
Moreover, the bandwidth of the calculated in-gap states increases as the size of the supercell decreases until the gapless ideal square-net in the 1x1x1 cell is reached, reproducing the experimentally observed trend. We estimate that the intermediate CDW phases of the potassium-doped surface have supercell sizes on the order of 10-20 unit cells.  
Because the center subcell for each model supercell is a perfect square ($y=0.5$) that is equivalent to that of \tLST, the size of the supercell effectively controls the fraction of unit cells that remains distorted. The increasing bandwidth of the in-gap states with doping thus indicates the gradual restoration of \textit{n} glide symmetry in the surface antimony layer, allowing for a continuous tuning of symmetry that does not seem to have a counterpart in the bulk materials, where \textit{n} glide symmetry in the antimony layer either exists or not.  
A similar doping dependence of CDW supercell size was observed previously in RESb\textsubscript{x}Te\textsubscript{2-x} systems~\cite{Dimasi1996a,Singha2021a,Lei2019a} but only in bulk samples, and the resulting changes to the band structure were not explored. 

Our study provides convincing evidence that electron doping causes a phase transition both in bulk (via chemical substitution) and on the surface (via chemical gating) of \LST\ through restoration of \textit{n} glide symmetry in the antimony layer.
While the transition is robust against the disorder that necessarily accompanies bulk chemical substitution, the electron doping at the surface is a cleaner, reversible, and versatile way to control the topological states. 
This supports the conclusion that the phase transition from \oLST\ to \tLST\ is electronically driven, and more generally that the electron concentration is the primary control parameter for the structural and topological transitions in \LST. 
We note, however, that the underlying electronic instability involves several factors. Fermi surface nesting is likely an important ingredient, given the geometry of the Fermi surface of \tLST, which features large parallel sheets that are destroyed by gap formation. 
In similar square-net materials, nesting has also been discussed as a component of CDW transitions~\cite{Dimasi1996a,Brouet2008a}, although other factors often dominate CDW transitions in such (quasi) 2D systems~\cite{Johannes2008a}. Here, the presence of negative frequencies in the calculated phonon spectrum for the structure of \tLST\ with no doping (x=1) indicates that, in addition to nesting, electron-phonon coupling does indeed play a role in the instability (see SM). Furthermore, other factors such as the density of states at the Fermi level and electron-electron interactions should also be taken into account in developing a comprehensive microscopic description of the instability in \LST.

\section{\label{sec:concl}Conclusion}
\noindent In summary, we have demonstrated an efficient way to manipulate the nodal loop topology in square-net materials. Exploiting the structural instabilities in \LST, we break the \textit{n}~glide symmetry that protects the nodal loop in the absence of SOC by changing the antimony concentration x. As evidenced in bulk materials with x=0.9 and x=1, the breaking of \textit{n} glide symmetry opens a gap of several hundred meV in the nodal loop, manifesting a drastic change of the topology.
The control parameter for this transition is the electron concentration of the square-net layer, which opens the possibility of control by gating. Here, we used chemical gating of the surface by potassium deposition to realize the reversible switching between the two topologically distinct phases. 
The full reversibility of the gap closing is a pre-condition for effective switching of topology between gapless and gapped nodal loop in devices. 
Taking our chemical gating study as a proof of concept, we expect that electrostatic gating can be used to control the topological phase transition in \LST\ thin films, allowing for wider doping regimes and more facile device integration.  
More generally, our doping-based approach is expected to apply to many other square-net materials, and therefore provides a direct path for identifying materials exhibiting topological phase transitions. Among those, we envision in particular materials with the nodal loop located near the Fermi level, which will be advantageous for applications based on, for example, electrical transport. In addition, many of the related square-net compounds order magnetically, which open the possibility of generating a Berry curvature-driven response by gapping out the nodal loop in the absence of time-reversal symmetry. Consequently, our findings define a new paradigm in the realization of functionalities that exploit transitions between topologically distinct states, a crucial step towards a family of devices based on the properties of topological materials. 

\section{\label{sec:acknowledge}Acknowledgements}
\noindent We thank Anette von der Handt (UBC Earth Sciences) for assistance with the wavelength dispersive X-ray spectroscopy measurements. 
This research was undertaken thanks in part to funding from the Max Planck-UBC-UTokyo Centre for Quantum Materials and the Canada First Research Excellence Fund, Quantum Materials and Future Technologies Program. This project is also funded by the Natural Sciences and Engineering Research Council of Canada (NSERC); the Canada Foundation for Innovation (CFI); the British Columbia Knowledge Development Fund (BCKDF); the Department of National Defence (DND); the Mitacs Accelerate Program;
the Moore EPiQS Program (A.D.); the Canada Research Chairs Program (A.D.); and the CIFAR Quantum Materials Program (A.D.).
This research is funded in part by a QuantEmX grant from ICAM and the Gordon and Betty Moore Foundation through Grant GBMF9616 to M.M. and H.-H.K. 
J.B. and H.-H.K. acknowledge the receipt of support from the CLSI Student Travel Support Program. 
J.W.S. was supported in part by a Provost's Research Fellowship from Farmingdale State College.
Use of the Canadian Light Source (Quantum Materials Spectroscopy Centre), a national research facility of the University of Saskatchewan, is supported by CFI, the NSERC, the National Research Council, the Canadian Institutes of Health Research, the Government of Saskatchewan, and the University of Saskatchewan.

\section{\label{sec:authors}Author contributions}
\noindent JB grew single crystals and characterized them. JWS performed single-crystal XRD and solved the crystal structures. JB, MM, and HHK conducted the ARPES experiments with help from MZ, SGor., and SZ.
JB analyzed the ARPES data with input from MM, HHK, and AD.
SGod. analyzed the core level spectra. JB and ISE performed DFT calculations. 
JB, MM, HHK, MO, AD, and MCA discussed the results. JB, MM, HHK, AD, and MCA wrote the paper, with contributions from all authors.

\section{\label{sec:competing}Competing interests}
\noindent The authors declare they have no competing interests.

\begin{figure*}[ht!]
	\centering
	\includegraphics[width=1\linewidth]{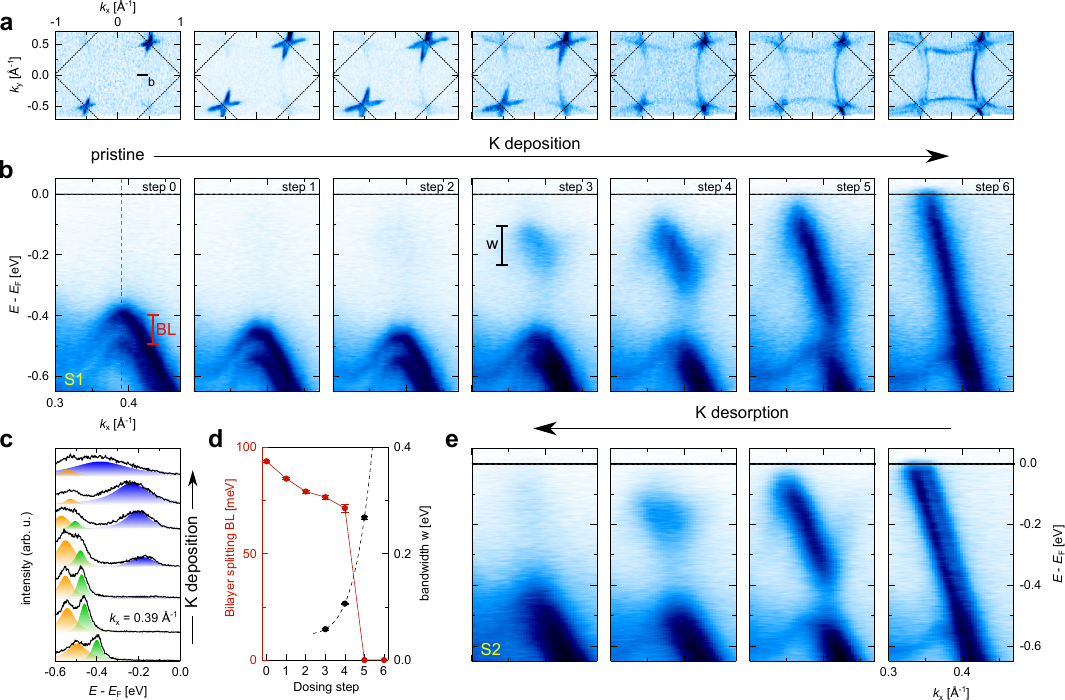}
	\caption{\label{fig3:kdepos} \textbf{Topological switching by \textit{in situ} control of crystalline symmetry.} 
    \textbf{a} Evolution of the Fermi surface of \oLST\ with stepwise potassium deposition. All data presented in this figure were acquired with LH polarization. After an initial increase in Fermi surface volume caused by the electron doping, the spectral weight redistributes until two concentric diamond shaped sheets with fourfold symmetry are formed in step 6.  
    \textbf{b} Evolution of the spectra along $\Gamma$-M (black line in \textbf{a}). The nodal loop gap closes via the formation of in-gap states that fill up the entire gap in step 6. Together with the fourfold symmetric Fermi surface, 
    this state resembles the electronic structure of bulk \tLST.
    \textbf{c} Representative energy distribution curves at $k_\mathrm{x}=0.39$~\AA\textsuperscript{-1} (black dashed line in \textbf{b}) for each dosing step and fits to pseudo-Voigt line shape. The upper band of the initial bilayer-split valence bands merges with the in-gap states to form the gapless nodal loop.
    \textbf{d} Reduction and disappearance of bilayer splitting with increasing potassium deposition, as extracted from panel \textbf{c}. At the same time, the bandwidth of the in-gap states increases to the full width of the initial gap in \oLST. The black dashed line is a guide to the eye. 
    Error bars are the standard deviations as obtained from the pseudo-Voigt fits shown in panel \textbf{c}.
    \textbf{e} Desorption of potassium transforms the gapless nodal loop back to the initial gapped state, indicating a fully reversible switching of topology, with the transition between gapless and gapped nodal loop mediated by manipulation of \textit{n} glide symmetry. Data in panels \textbf{a}-\textbf{d} were measured on sample S1, and in panel \textbf{e} on sample S2.} 
\end{figure*}

\begin{figure*}[ht!]
	\centering
	\includegraphics[width=0.5\linewidth]{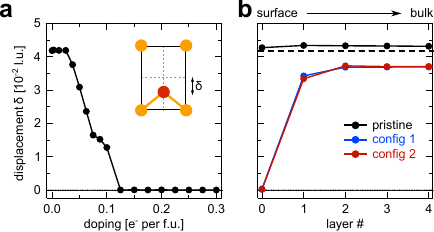}
	\caption{\label{fig4:dft} \textbf{Doping-dependent breaking of \textit{n} glide symmetry from DFT.}
    \textbf{a} Evolution of the center antimony displacement $\delta$ with pure electron doping in bulk \oLST. The structure for each doping level was relaxed. The displacement becomes 0 and \textit{n} glide symmetry is restored for doping levels larger than 0.13 electron per formula unit, indicating that electron doping alone can explain the phase transition from \oLST\ to \tLST\ in bulk. The inset shows the definition of the displacement of the center Sb atom. 
    \textbf{b} Center antimony displacement as a function of layer number for three relaxed 5-unit cell slabs of \oLST\ (pristine and K adatoms in two different geometries, see SM for details). For both potassium-covered slabs, \textit{n} glide symmetry is restored in the first antimony layer, while deeper layer remain heavily distorted, highlighting that the main effect of potassium is electron doping. 
    }
\end{figure*}

\begin{figure*}[ht!]
	\centering
	\includegraphics[width=0.5\linewidth]{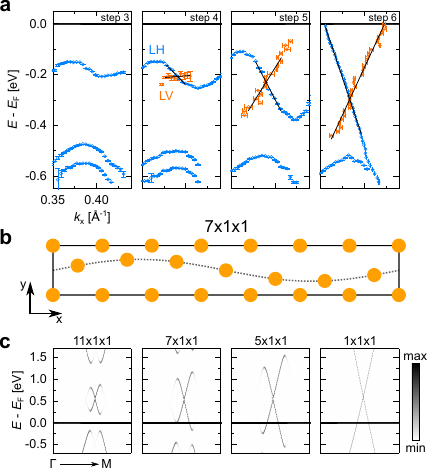}
	\caption{\label{fig5:cdw} \textbf{Continuous tuning of \textit{n} glide symmetry.}
    \textbf{a} Band dispersions obtained from fits of spectra along $\Gamma$-M acquired with LH (blue symbols, same spectra as in Fig.~\ref{fig3:kdepos}b) and LV polarization (orange symbols), for potassium dosing steps 3-6. 
    The spectra are shown in the SM.
    Black lines are linear fits around the band crossings of the in-gap states. Steps 4 and 5 exhibit the coexistence of Dirac cone and nodal loop gap, which cannot be explained by the model in Fig.~\ref{fig1:square}. The observation of back folded bands in the in-gap states in steps 3-5 indicates CDW phases.
    Error bars are the standard deviations as obtained from pseudo-Voigt fits to the energy and momentum distribution curves of the spectra.
    \textbf{b} Structure of the proposed CDW model that qualitatively describes the data in panel \textbf{a}. Shown here is a 7x1x1 supercell of monolayer antimony with sinusoidal variation of the center antimony's \textit{y} coordinate. The amplitude of the variation is exaggerated by factor of 4 for visualization.
    \textbf{c} Band structures of the CDW supercells with varying size of supercell, after unfolding and projection onto \textit{p}\textsubscript{x,y} states. In-gap states show band folding similar to the experimental data in panel \textbf{a}. With decreasing size of the CDW supercell the bandwidth of the in-gap states increases, replicating the trend in the experimental data. Note that the 1x1x1 ``supercell'' corresponds to pristine \tLST. Fermi level shifts observed in the data in panel \textbf{a} are ignored in the model.
    }
\end{figure*}



%

\clearpage

\section{\label{sec:meth}Methods}
\noindent \textbf{Single crystal growth.}
Single crystals of \oLST\ were grown using the liquid transport method, inspired by Ref.~\cite{Yan2017a}. Elements were weighed in a 1:15:2-x ratio (La:Sb:Te) and sealed in a quartz glass tube under 0.3~bar argon. The mixture was then heated to 910~$^\circ$C and kept in a horizontal gradient from 910~$^\circ$C to 770~$^\circ$C for 10-14~days before cooling to room temperature at a rate of 50~K/h. Crystals were separated from the solidified flux after heating the mixture to 700~$^\circ$C for 20~min and then spinning.
\tLST\ was grown by chemical vapor transport. Elements were weighed in a 1:5:1 ratio (La:Sb:Te) with an addition of 100~mg iodine and sealed in a quartz glass tube under vacuum. The mixture was pre-reacted at 800~$^\circ$C for one week and then subjected to a temperature gradient from 910~$^\circ$C to 800~$^\circ$C for two~weeks with the crystals forming in the hot end. 

\noindent \textbf{Characterization.}
The phase purity of the crystals was checked with powder XRD. Patterns on ground single crystals were acquired with a Bruker D8 using Cu K$\alpha_1$ radiation. The crystal structures of \oLST\ and \tLST\ were solved from single crystal XRD. Data were acquired with a Rigaku Xta{\sc lab} mini II using Mo K$\alpha$ radiation and the structures were refined by a standard charge-flipping method. Crystal structures were visualized with VESTA~\cite{Momma2011a}.
The composition of the crystals was determined by wavelength dispersive X-ray spectroscopy. 

\noindent\textbf{ARPES measurements.}
Measurements were conducted at the Quantum Materials Spectroscopy Center (QMSC) beamline at the Canadian Light Source. Sample were cleaved and measured at \textit{T}=20~K and a base pressure of $6\cdot10^{-11}$~torr using 87~eV photons, unless mentioned otherwise. The angular resolution was 0.1~° and the energy resolution was about~30~meV. \textit{In situ} potassium deposition was performed in six steps for 5~s, 5~s, 5~s, 15~s, 30~s, and 60~s (steps 1-6 in Fig.~\ref{fig3:kdepos}a) at \textit{T}=20~K using a getter source from SAES. For potassium desorption experiments, the sample was transferred \textit{in situ} into a heating stage inside the ultra-high vacuum chamber and heated in three steps (15~min at 100~°C, 15~min at 120~°C, and 45~min at 130~°C). 
The pressure during this procedure did not exceed $4\cdot10^{-10}$~torr. After each heating step, the sample was transferred back into the cryostat and cooled to 20~K, followed by acquisition of the spectra. 

\noindent \textbf{DFT calculations.}
Electronic structure calculations were performed within the framework of DFT as implemented in the package WIEN2k~\cite{Blaha2020a,Blaha2018a}. The generalized gradient approximation with the Perdew-Burke-Ernzerhof (PBE) parametrization~\cite{Perdew1996a} was used. The basis set size was set to R\textsubscript{mt}K\textsubscript{max}=8.5 and the Brillouin zone was sampled with 6000 k~points. The results of these calculations were cross-checked with Quantum Espresso~\cite{Giannozzi2009a,Giannozzi2017a}. 
Furthermore, supercell calculations presented in Fig.~\ref{fig5:cdw}c were performed using Quantum Espresso. The plane wave expansion was cut off at 60~Ryd and the Brillouin zone was sampled with about 1~k point per 0.1 \AA$^{-1}$. The bands were unfolded into the Brillouin zone of the 1x1 cell using code developed in Ref.~\cite{Pacile2021a}.
DFT calculations shown in Fig.~\ref{fig4:dft} were performed with the pseudopotential code VASP \cite{Kresse1996a,Kresse1996b,Kresse1999a} using the projector augmented wave (PAW) method~\cite{Bloechl1994a}. The PBE functional \cite{Perdew1996a} was used for the exchange-correlation energy. The kinetic energy cutoff was 600~eV. The bulk Brillouin zone was sampled using a 16x16x4 k-grid, while slab calculations were carried out with a 12x12x1 grid. We used slabs of 5 unit cell thickness. All ionic positions were relaxed until the norms of all forces were less than 0.01~eV/\AA.
Phonon dispersions were calculated with Phonopy~\cite{Togo2023a} using 3x3x2 supercells containing 108~atoms. The total energy calculations were performed using VASP with a k~mesh of 7x7x4 and a cutoff energy of 700~eV.

\section{\label{sec:data}Data availability}
\noindent All data needed to evaluate the conclusions in the paper are present in the paper. The raw ARPES data acquired in this study have been deposited in the Zenodo database under the digital object identifier \url{https://doi.org/10.5281/zenodo.17059971}~\cite{Bannies2025b}. All other data are available from the corresponding authors upon reasonable request. 

\end{document}